\documentclass[
   preprint
,
 superscriptaddress,
 amsmath,amssymb,
 aps ,prl
]{revtex4-1}

\usepackage{graphicx}
\usepackage{dcolumn}
\usepackage{bm}
\usepackage{color}
\usepackage{soul}

\begin{document}

\preprint{APS/123-QED}

\title{Ion specificity modulated inhomogeneous interfacial flow inhibits bubble coalescence in electrolyte solutions}


\author{Bo Liu}
 \affiliation{%
 Department of Chemical and Materials Engineering, University of Alberta, Edmonton,T6G 1H9, Canada\\
 }%
\author{Rogerio Manica}%
\affiliation{%
 Department of Chemical and Materials Engineering, University of Alberta, Edmonton,T6G 1H9, Canada\\
}%


\author{Zhenghe Xu}
\affiliation{%
 Department of Chemical and Materials Engineering, University of Alberta, Edmonton,T6G 1H9, Canada\\
}%
\affiliation{%
	Department of Materials Science and Engineering, Southern University of Science and Technology, Shenzhen, 518055, China\\
}%
\author{Qingxia Liu}
\email{qingxia2@ualberta.ca}
\affiliation{%
 Department of Chemical and Materials Engineering, University of Alberta, Edmonton,T6G 1H9, Canada\\
}%


\date{\today}

\begin{abstract}

Inhibition of bubble coalescence in electrolyte solutions enables the formation of oceanic whitecaps~\cite{oceanwave_bubble} and affects the heat and mass transfer in many bubble related engineering processes. For different electrolytes, the ability to inhibit bubble coalescence correlates to the ion specificity at the air water interface at an abnormal cation-anion pair relationship~\cite{craig1993effect,craig2011hydration}, rather than the typically expected cation or anion series that was widely reported in atmospheric, bio- and chemical processes~\cite{jungwirth_specific_ion2006,jungwirth2014beyondhofmeister,pegram2007hofmeister}


Here we show that the inhomogeneous interfacial flow, at a different electrolyte concentration from the solution because of the surface specificity of both cation and anion~\cite{levin_langmuir,levin_prl_ionsatinterface}, contributes to the bubble coalescence inhibition behavior in electrolyte solutions. The interfacial flow, achieved with the mobile air-water interface~\cite{liu2019coalescence}, contributes to the continuous change of electrolyte concentration within the liquid film formed between two colliding bubbles, thereby resulting in a concentration gradient of electrolytes between the thin film and the bulk solution. The electrolyte concentration gradient, hence surface tension gradient, becomes significant to resist film thinning when the film thickness reaches tens of nanometers. The retarded film thinning between two bubbles and delayed bubble coalescence were experimentally captured by high-speed interferometry and quantitatively explained by the proposed electrolyte transportation model. This finding clearly highlights the coupled effect of interfacial flow and ion specificity, and shows important implications for improved understanding of ocean waves~\cite{oceanwave_bubble,bird2010daughterbubble}, for future development of colloidal science in high concentration electrolyte solutions that is critical for biology~\cite{ninham2017surface,jungwirth2014beyondhofmeister}, and benefits many applied fields like water splitting~\cite{bubble_electrochemical} and mineral extraction~\cite{yoon1993microbubble}.

\end{abstract}

\maketitle
 
\begin{figure}[htb]
	\includegraphics[width=0.95\textwidth]{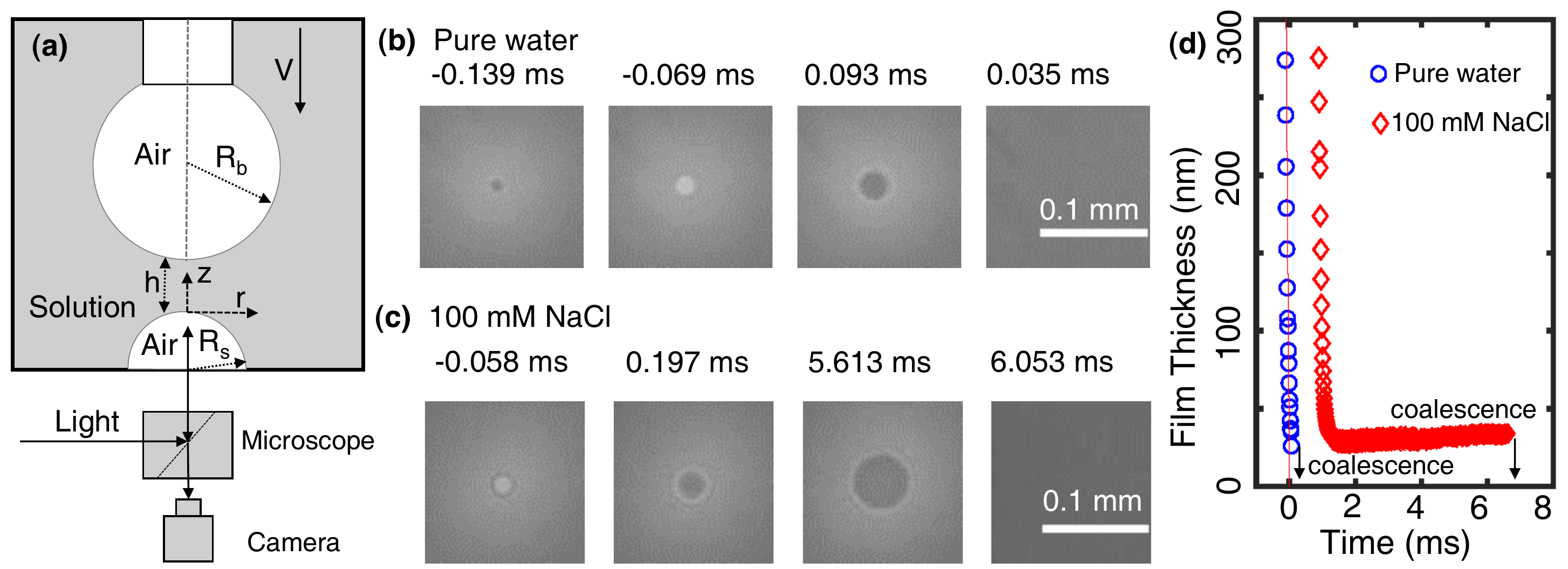}
	\caption{\label{fig:sche}
		(a) A sketch of the experimental setup. A bubble of radius $R_b$ = 1 mm was held at the glass capillary orifice, whereas another bubble of radius $R_s$ was immobilized on a hydrophobic silica surface. The capillary was driven downward with velocity $V$ by a speaker diaphragm to allow bubble collision. A high-speed camera connected to an inverted microscope was used to record the interference fringes. Snapshots of the fringes in a time sequence obtained between two colliding bubbles in (b) pure water ($R_s$ = 0.41 mm) and (c) 100 mM NaCl solutions ($R_s$ = 0.42 mm). By analyzing the interferometric images, (d) the film thickness evolution was obtained, NaCl curve is shifted for clarity.
	}
	
\end{figure}

In the experiment, two bubbles submerged in an electrolyte solution are brought together to study their collision and coalescence behavior(Fig.\ref{fig:sche}a). A top bubble (radius $R_b$ = 1.00 mm, generated and held at a capillary orifice) was driven towards a bottom bubble (radius $R_s$ between 0.10 and 0.80 mm, immobilized on a transparent hydrophobic silica surface with contact angle $\sim$100$^{\circ}$) at a controlled velocity of around 3 mm/s, unless specified otherwise. A light beam was sent in through an inverted microscope and the reflection of light from both bubble surfaces interfered with each other, changing the light intensity that was recorded by a high speed camera at 86400 frames per second. By analyzing the light intensity, we were able to extract the spatiotemporal evolution of the film thickness $h$ between two bubbles at nanometer resolution.


Figs. 1b and 1c show the evolution of interferometric fringes between two colliding bubbles in pure water and in 100 mM NaCl solution, respectively. In pure water, the light intensity evolved rapidly, reflecting the fast liquid drainage. The film ruptured at a critical thickness in the range of 20 to 30 nanometers when a dark color was observed at the center. In 100 mM NaCl solution, the film thinning process could be divided into two distinct stages. The initial stage was similar to that in pure water featured by the small film width and rapid light intensity evolution. Later, the thinning process was suddenly `arrested' when a dark color appeared and started to expand.  

By analyzing the light intensity at the film center, the time evolution of the film thickness can be obtained and is presented in Fig. 1d.  For comparison, we define the time `$t=0$' as the time when the surfaces  would have touched if they were undeformable (e.g., rigid spheres). This definition sets a reference time to measure how long the coalescence was delayed (termed the `coalescence time') in different solutions. In pure water, the film thinned with negligible delay enabling almost instantaneous coalescence with time close to 0 ms (see Fig. 1d), indicating that the air-water interfaces had fully mobile boundary conditions, a feature that was recently confirmed using the same technique~\cite{liu2019coalescence}. In NaCl solution, the film thinning was almost identical to pure water until the thickness reached around 30 nm, after that the film thinning was hindered for around 6 ms before rupture. 

The retardation of film thinning at tens of nanometers is unexpected from currently known mechanisms. Surface forces, such as the repulsive electrical double layer, affect the film thinning at a similar thickness, but they should stabilize the film for much longer rather than just for milliseconds. Delay of film thinning can also be observed in surfactant solutions, but at a much larger thickness of a few microns and for longer time scales under similar experimental conditions~\cite{liu2019coalescence}. To better understand this retardation feature that determines the coalescence time, a systematic investigation was conducted by taking into consideration of bubble size, electrolyte concentration and type.

Two-stage thinning of the liquid film was consistently observed in NaCl solutions, but the coalescence time varied significantly with bubble size and electrolyte concentration (Fig. 2a). By increasing NaCl concentration, the coalescence time increased significantly from sub-milliseconds in 50 mM to $\sim$100 ms in 250 mM. Bubble size was also shown to increase the coalescence time by one or two orders of magnitude by using larger bubbles. By increasing NaCl concentration in the same range, the reported coalescence probability in a bubble column was shown to decrease from 100\% to almost 0\%~\cite{Craig1993}. 

\begin{figure}[htb]
	\includegraphics[width=0.95\textwidth]{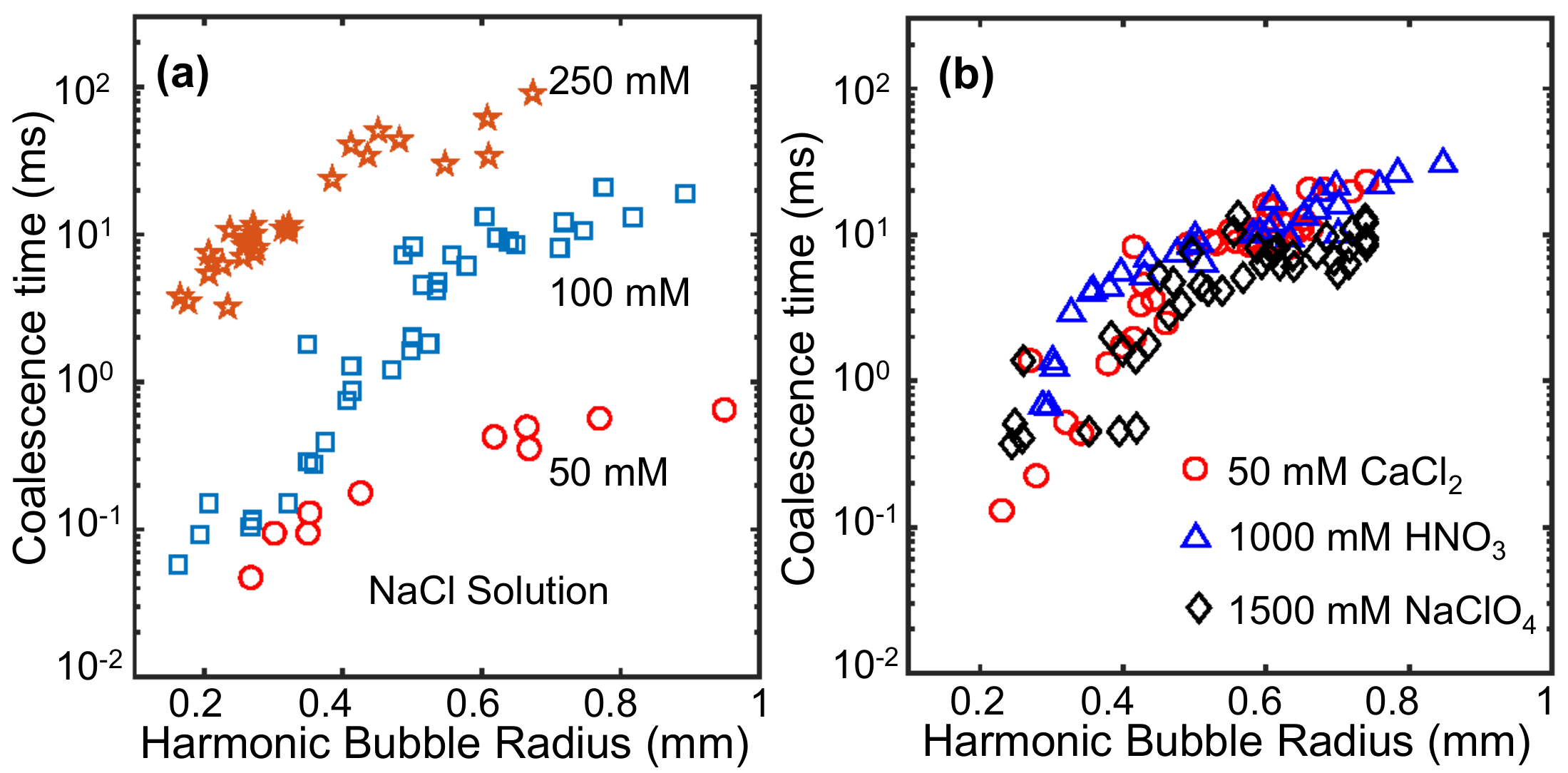}
	\caption{\label{fig:light}
		The coalescence time as a function of harmonic mean radius $R_H=2R_sR_b/(R_s+R_b)$, in (a) NaCl solutions at different concentrations; and (b) different electrolyte solutions at different concentrations.
	}
\end{figure}

The film thinning features were very consistent and reproducible in various electrolyte solutions such as KCl, HCl, NaOH, HNO$_3$ ,H$_2$SO$_4$, NaClO$_4$, CaCl$_2$, etc. The ability to influence the retardation stage, however, varied significantly with electrolyte type. As shown in Fig. 2b, 50 mM CaCl$_2$ provided similar coalescence times as 100 mM NaCl. In contrast,  it requires 1000 mM HNO$_3$ and 1500 mM NaClO$_4$ to achieve the similar coalescence time. These results are consistent with reported data that CaCl$_2$ could inhibit bubble coalescence at a lower concentration~\cite{marrucci1967coalescence,craig1993effect}, whereas NaClO$_4$ required a much higher concentration of up to 1500 mM to affect the bubble coalescence~\cite{christenson2008electrolytes}.

Our results so far have clearly demonstrated that electrolytes affect bubble coalescence by delaying film thinning for milliseconds. We note that in a dynamic bubbling system, such as a bubble column, the colliding bubbles are estimated to interact for a limited time in the range of a few milliseconds~\cite{kirkpatrick1974influence,liu2019coalescencejpcl}. Bubbles would bounce away if the required coalescence time is longer than the interaction time. Therefore, delaying the film thinning between two bubbles for milliseconds is crucial to determine the coalescence probability in dynamic environment. Its dependence on electrolyte type, concentration and bubble size that were reported in the literature using various techniques can now be universally explained by this finding. For example, non-coalescence of bubbles in the ocean wave~\cite{oceanwave_bubble} now can be explained by the salinity of around 3.5 wt \% ($\sim$600 mM NaCl) in sea water that can significantly retard the film thinning. 


The direct observation (in time and space) of the thin film separating the bubbles provides an excellent opportunity to theoretically explain the coalescence inhibition mechanism by developing a suitable model that predicts the main experimental features. The model should be able to capture the following (see also Fig. 3):

\begin{figure}[htb]
	\includegraphics[width=0.95\textwidth]{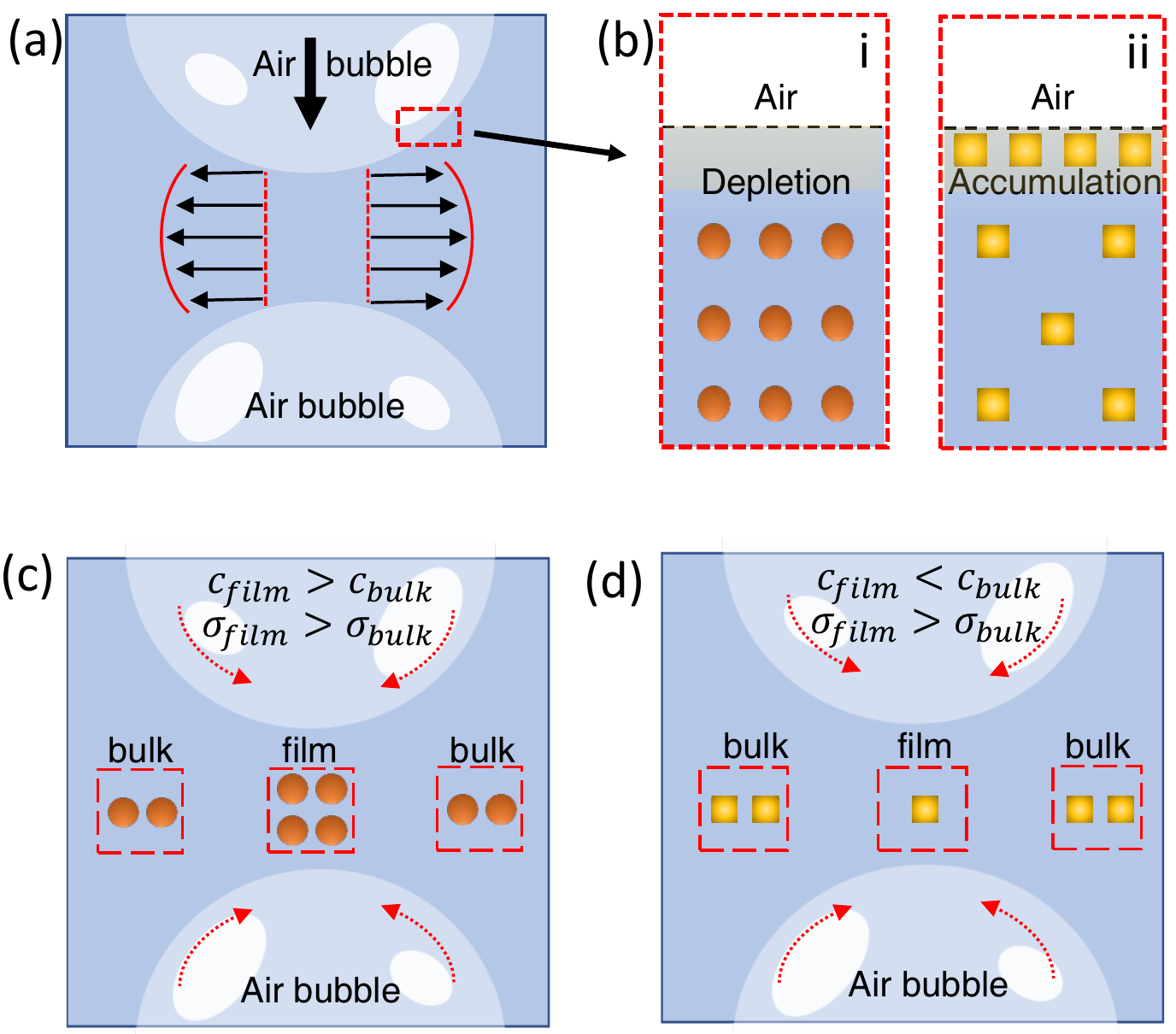}
	\caption{\label{fig:com}
		Schematics for (a) the fluid flow within the thin liquid film with mobile air-water interfaces; (b) the electrolyte distribution close to the Gibbs dividing plane for the cases of: (i) exclusion (ii) accumulation of ions at the interface. In the film thinning process, (c) surface exclusion results in an increased electrolyte concentration at the film region; (d) surface accumulation results in a decreased electrolyte concentration in the film. Both contribute to a higher surface tension at the film center.
	}
\end{figure}

(1) At the initial approach stage, the air-water interfaces should be fully or partially mobile featured by a surface velocity $U$ (Fig. 3a). 

(2) At the retardation stage, the ability to affect the film thinning, hence the coalescence time, varies significantly with electrolytes. The results in Fig. 2 clearly demonstrate the inhibition effect decreases following the order:  CaCl$_2$$>$NaCl$>$HNO$_3$$>$NaClO$_4$, following a similar trend in which these electrolytes affect the air-water interfacial tension.

According to the experimental results, the change of interfacial tension as a function of electrolyte concentration obeys an approximately linear relationship~\cite{pegram2006partitioning,pegram2007hofmeister}, whose slope varies with electrolytes. For example, the interfacial tension can be changed by $\sim$1.7 mN/m with the addition of 1000 mM NaCl, while the change at the same concentration is approximately 3.3 mN/m for CaCl$_2$, -0.8 mN/m for HNO$_3$ and 0.2 mN/m for NaClO$_4$~\cite{pegram2006partitioning,pegram2007hofmeister,ionhydration_coalescence_cacl2_gascontent}. 



Electrolytes affect the interfacial tension by their exclusion or accumulation  from the interfacial region~\cite{pegram2006partitioning, levin_prl_ionsatinterface,levin_langmuir}, features that are schematically  shown in Figs. 3b (i) and (ii), respectively. For example, surface excluded electrolytes (e.g., NaCl) tend to increase the surface tension, while surface accumulated electrolytes (e.g., HNO$_3$) reduce it. 
Despite the inhomogeneous distribution of ions within the interfacial region, the electrolyte partition can be quantified   
 from experimental results by the Gibbs definition of surface excess $\Gamma=-\frac{c}{RT(1+\epsilon_{\pm})}\frac{d\sigma}{dc}$, where $R$ is the gas constant, $T$ is temperature, $\epsilon_{\pm}$ is a non-ideality correction term that described in the supporting information~\cite{ionhydration_coalescence_cacl2_gascontent,marrucci1967coalescence}. Using this definition, the depth of the inhomogeneous interfacial region is estimated to be a few angstroms.  
The electrolyte  concentration at the interfacial region is not only related to the individual surface specificity of cation and anion, but also influenced by the interplay between the ions possibly through the electrostatic interaction~\cite{levin2009ions,levin_langmuir}. In other words, there is a cation-anion pair effect on interfacial tension in electrolyte solutions.  

To resist the film thinning, an interfacial tension gradient $\sigma_{film}>\sigma_{bulk}$ is required to balance the shear stress~\cite{liu2019coalescence}. In surfactant solutions, it can be achieved purely by the surface convection of surfactant~\cite{liu2019coalescence,Chan2011a,liu2019coalescencejpcl}. However, this mechanism is not applicable in electrolyte solutions considering the fast equilibrium of electrolytes between the solution and the interfacial region in pico-seconds (need cite).
Instead, it can be achieved by an electrolyte concentration gradient between the film solution ($c_{film}$) and the bulk soluton ($c_{bulk}$), hence the required interfacial tension gradient. The concentration gradient is possibly enabled by the electrolyte flow from the inhomogeneous interfacial region.

In the film thinning process, water and electrolyte are flowing outward from the film solution and through the interfacial region at a different electrolyte concentration. For surface excluded electrolytes, a lower electrolyte concentration at the interfacial depletion region enables an additional amount of water to flow out. As a consequence, the electrolyte concentration increases continuously inside the liquid film (see Fig. 3c), resulting in a higher electrolyte concentration in the film solution than in the bulk solution ($c_{film}$$>c_{bulk}$), hence $\sigma_{film}>\sigma_{bulk}$ that resists film thinning.
For surface accumulation electrolytes, the film concentration tends to decrease ($c_{film}$$<c_{bulk}$, see Fig. 3d). However, the surface tension gradient remains in the same direction since a lower concentration results in a higher interfacial tension ($\sigma_{film}>\sigma_{bulk}$). 

To quantify the transport of electrolyte in the thin film, we developed a model that considers the solution advection in the film and at the surfaces, which is written as:
\begin{eqnarray}
	\label{massbalance}
	\begin{aligned}
		&  h\left[ \frac{\partial c}{\partial t}+\left(U+\frac{h^2}{12\mu}\frac{\partial p}{\partial r} \right)\frac{\partial c}{\partial r}-\frac{D}{r}\frac{\partial}{\partial r}\left(r\frac{\partial c}{\partial r}\right) \right] \\
		&\ +2 \left[\frac{\partial \Gamma}{\partial t}+\frac{1}{r}\frac{\partial(r\Gamma U)}{\partial r}-\frac{D_s}{r}\frac{\partial}{\partial r}\left(r\frac{\partial \Gamma}{\partial r}\right)\right] 
		=0
	\end{aligned}
\end{eqnarray} 
where the first term represents the electrolyte flow inside the film and the second term represents the same process at the bubble surfaces. In this equation, $r$ is the radial coordinate, $\mu$ the liquid viscosity, $D$ and $D_s$ are the diffusion coefficients in the bulk and at the surface, respectively. Obviously, the importance of surface transport (second term) becomes more significant as the film thickness $h$ decreases. 


Starting with an initial concentration $c=c_o$, Eq.~\ref{massbalance} is rearranged 
and solved numerically coupled with the lubrication equation for the liquid flow inside the film~\cite{davis1989lubrication,liu2019coalescence}, and the Young-Laplace equation for surface deformation~\cite{Chan2011a,liu2019coalescence}. The complete theoretical model and details of the numerical technique are provided in the supporting information.

Without any fitting parameters, notable agreement between model predictions and experimental results is shown in Fig. 4a for the evolution of the film thickness as a function of time. As consistently predicted and observed in different concentrations, electrolyte shows little effect on the initial rapid film thinning stage. In the later stage, a significant delay on film thinning was predicted at a thickness of tens of nanometers that varied slightly with concentration. 


\begin{figure}[htb]
	\includegraphics[width=0.85\textwidth]{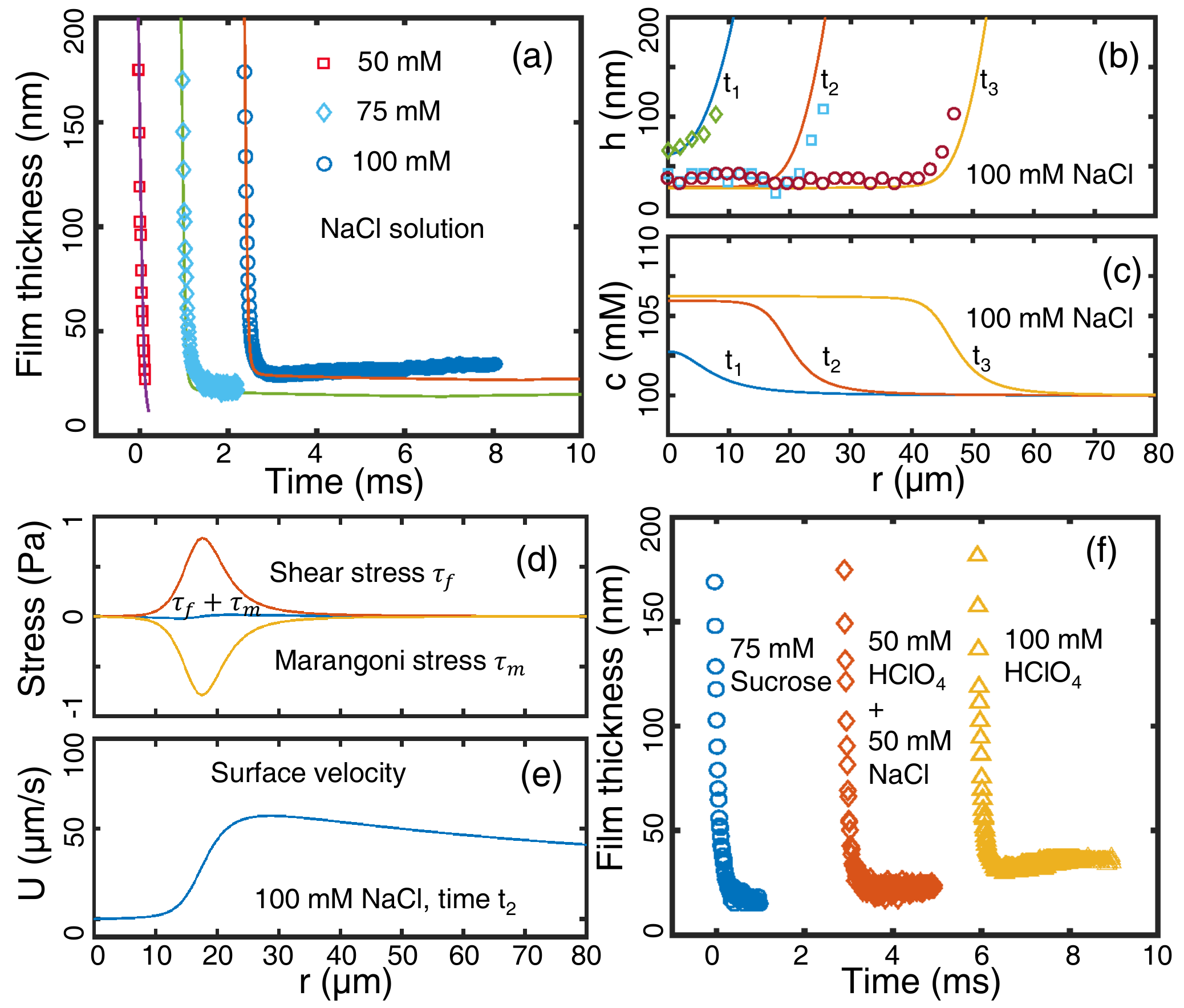}
	\caption{\label{fig:last}
		(a) Comparison between theory and experimental observations of the film thickness evolution at the film center; three concentrations are included: $c_o$ = 50 mM, $R_s$ = 0.19 mm,  $c_o$ = 75 mM, $R_s$ = 0.42 mm, $c_o$ = 100 mM, $R_s$ = 0.42 mm; curves are shifted for clarity.   (b) The lateral film thickness at three times ($t_1$=-0.01 ms, $t_2$=1.13 ms, $t_3$=4.59 ms) in 100 mM NaCl solution,$R_s$ = 0.47 mm; (c) The NaCl concentration distribution within the thin liquid film; At time $t_2$, (d) the fluid shear stress and Marangoni stress distribution and (e) the surface velocity distribution within the thin liquid film. (f) The film thinning behavior in sucrose solution with $R_s$ = 0.43 mm, in the mixture of NaCl and NaClO$_4$ with $R_s$ =0.45 mm, and in HClO$_4$ solution with $R_s$ = 0.49 mm, curves are shifted for clarity.
	}
\end{figure}

The slowdown of thin film drainage and the flattening of the curved bubble surfaces are also successfully captured by the model (see comparison in Fig. 4b). The flattening represents that the hydrodynamic pressure becomes closer to the bubble Laplace pressure. The former is negligible with fully mobile air-water interfaces, and that is the reason the surfaces barely deform at the initial stage. Surface deformation at the retardation stage is a consequence of the reduced interface mobility. These results indicate that the uneven transport of electrolyte  affects the air-water interface mobility significantly at the retardation stage, but not at the initial stage when film is thicker, in agreement with the finding that surface transport becomes more important at small film thickness. Agreement between theory and experiment validates the model performance on predicting the film thickness evolution, the transition of boundary conditions and the pressure distribution. The model can now be used to improve our understanding of the two-stage film thinning due to electrolytes. 
  
The origin of the two-stage thinning behavior is the formation of an electrolyte concentration gradient within the thin liquid film. Fig. 4c shows the predicted concentration distribution inside the film region at different times corresponding to 4b. The initial concentration of 100 mM had negligible variation until the film thickness reached around 50 nm. Afterwards, the concentration at the film center started to rise and reached the peak of $\sim$105 mM accompanied by the flattening of the surfaces. In the radial direction, a similar increase-peak relationship occurred with the flattening of the thin film. Once the flattened film was reached, the concentration at the flatten region stayed mostly constant. 
  
The predicted variation of the electrolyte concentration was small ($\sim$5\%) (Fig. 4c), but its effect on film thinning was remarkable. The surface tension gradient ($d\sigma/dr$) arising from the concentration gradient ($dc/dr$) was around $10^{-5}$ N/m. But it changed quickly along the $\sim$10 $\mu$m transition region giving rise to the Marangoni stress $\tau_m=d\sigma/dr$ of $\sim$1 Pa that was sufficient to counterbalance the fluid shear stress $\tau_f$. Fig. 4d depicts the negative (inward) Marangoni stress plus the positive (outward) shear stress giving rise to an overall stress close to 0. The mirror like distribution of the stresses is a consequence of the fluid shear stress originating the concentration distribution, which in turn creates the Marangoni stress. This explains why the electrolyte (NaCl) with minimal surface tension effect 
can influence the film thinning in such a dramatic way. The overall stress profile enabled a surface velocity close to 0 at the flattened region, but much larger at the curved region (see Fig. 4e). The surface velocity of the two different regions led to the `arrest’ at the flattened region, and the continuous approach of the curved region, resulting in the expansion of the flat film. This flattening of the film is very different from a system with surfactants, where the film tends to be curved~\cite{liu2019coalescence,Chan2011a}. 

The slowdown thickness is the most important parameter that determines the measured coalescence time. Reducing this value by a few nanometers allows the film to be ruptured by the attractive van der Waals force, as observed in 50 mM NaCl solution. A simplified scaling analysis of the developed model links the slowdown thickness $h_s$ with experimental properties as $h_s\sim  \sqrt{(d\sigma/dc)^2(c_oR_H/\sigma RT)} $. This relationship clearly highlights the key experimental revealed parameters including electrolyte concentration $c_o$, electrolyte type through $d\sigma/dc$, and bubble size $R_H$.  


Using the example of NaCl, delayed film thinning is explained by the concentration increase within the film region that contributes to a larger interfacial tension. For a  surface enriching electrolyte such as HClO$_4$, the model predicts that the concentration decreases as the surfaces approach, but once again inward Marangoni stress that resist the film thinning (see supporting information). Therefore, the coalescence inhibition is explained using the same principle for both the surface exclusion and accumulation electrolytes.
 
 
 Given the above result, what will happen in the case of a mixture of 1:1 electrolyte of NaCl and HClO$_4$ that has a similar interfacial tension as pure water~\cite{henry2007ion}? 
 The surface excess of two electrolytes are relatively independent from each other (see supporting information). 
 Using this information, the model predicts an increased concentration of NaCl, and a decreased concentration of HClO$_4$ at the film solution, both increasing the Marangoni stress that resists film thinning. This analysis suggests that the inhibition effect of electrolytes is additive, agree with the experimental results on film retardation (see Fig. 4f) and coalescence time~\cite{henry2007ion}.
 Furthermore, it is interesting that the mixture inhibits bubble coalescence following the addition effect of NaCl and HClO$_4$, rather than the equivalent pair of NaClO$_4$ and HCl, the latter predicting a much weaker coalescence inhibition. This finding indicates the possible contribution from ion-ion correlation at the interfacial region~\cite{PNAS_ion_ion_correlation_interface}.


 
 
Applying this new knowledge allows us to explain many unexplained results obtained in the past five decades and extend the applications to a wider scope. For example, there is a very successful cation-anion pair empirical relationship that predicts whether one electrolyte would be able to inhibit bubble coalescence at the concentration of up to 0.5 M~\cite{craig1993effect,henry2007ion,ninham2017surface,craig2011hydration}. This relationship highlights the effect of ion pairs, rather than single ions that was emphasized by the famous Hofmeister series on protein precipitation. Our new understanding suggests that the ion pair relationship on coalescence inhibition arises from the ion partition behavior, the same physical origin of the Hofmeister series.   While direct ion-surface interaction is emphasized in protein precipitation~\cite{levin2009ions,pegram2007hofmeister}, the coalescence inhibition relies on the quantitative ion distribution at the interfacial region, i.e., surface excess, which is jointly determined by the cation-anion pair rather than single ions. 

 Here, we revealed a new mechanism for electrolytes to inhibit bubble coalescence under dynamic conditions. Similar behavior should occur for solutes more broadly as long as they have interfacial affinity, which is almost always true for all solutes considering the sharp transition of solvent environment near the gas-liquid interface, including the obvious changes on dielectric environment, the solvent structure and the solute-solvent interaction energy. Therefore, this new mechanism is not limited to electrolytes in aqueous solution, but also valid for electrolyte in non-aqueous solutions, and non-ionic solutes in aqueous or non-aqueous solutions~\cite{henry_non_aqueous,craig1993effect}. As expected, the two-stage film thinning behavior is also observed in systems such as sucrose in water (see Fig. 4f). 
 
 The early research and understanding of surfactants on bubble coalescence allowed improvements of foams, shampoos and detergent that facilitates our daily lives. The current understanding with barely-surface-active solutes can be beneficial to understand the bubbling behavior widely observed in nature and in engineering processes. Furthermore, this mechanism should be valid in a very fast time scale of $\mu$s or ns, much faster than the surfactant adsorption in ms. This unique feature will enable barely-surface-active solutes to be used to manipulate bubble coalescence in ultrafast processes, such as the generation of bubbles in a cavitation tube and the detachment of bubbles from a water splitting electrode that is hindered by the consistent coalescence of bubbles. Furthermore, realizing that the abnormal cation-anion pair relation has the same physical origin as the Hofmeister series will help to better apply the information obtained with bubbles to much complicated biological systems ~\cite{ninham2017surface,jungwirth2014beyondhofmeister}, in which the effect of co-ions and the solute additivity effect are attracting attention recently~\cite{jacs_salt_noadditive,jungwirth2014beyondhofmeister}. Finally, importance of the liquid flow on the interfacial chemistry, which has been reported on solid-water interface~\cite{liquidflow_solid_science}, also shows profound effect at the liquid/liquid interface that deserve further investigation.

\begin{acknowledgments}
 We thank the suggestions from Dr. Evert Klaseboer (IHPC, Singapore).	We acknowledge the financial support of the Natural Science and Engineering Research Council of Canada and the Canadian Centre for Clean Coal/Carbon and Mineral Processing Technology (C5MPT).  
\end{acknowledgments}

  \bibliography{salt_ref}

\begin{thebibliography}{29}%
\makeatletter
\providecommand \@ifxundefined [1]{%
 \@ifx{#1\undefined}
}%
\providecommand \@ifnum [1]{%
 \ifnum #1\expandafter \@firstoftwo
 \else \expandafter \@secondoftwo
 \fi
}%
\providecommand \@ifx [1]{%
 \ifx #1\expandafter \@firstoftwo
 \else \expandafter \@secondoftwo
 \fi
}%
\providecommand \natexlab [1]{#1}%
\providecommand \enquote  [1]{``#1''}%
\providecommand \bibnamefont  [1]{#1}%
\providecommand \bibfnamefont [1]{#1}%
\providecommand \citenamefont [1]{#1}%
\providecommand \href@noop [0]{\@secondoftwo}%
\providecommand \href [0]{\begingroup \@sanitize@url \@href}%
\providecommand \@href[1]{\@@startlink{#1}\@@href}%
\providecommand \@@href[1]{\endgroup#1\@@endlink}%
\providecommand \@sanitize@url [0]{\catcode `\\12\catcode `\$12\catcode
  `\&12\catcode `\#12\catcode `\^12\catcode `\_12\catcode `\%12\relax}%
\providecommand \@@startlink[1]{}%
\providecommand \@@endlink[0]{}%
\providecommand \url  [0]{\begingroup\@sanitize@url \@url }%
\providecommand \@url [1]{\endgroup\@href {#1}{\urlprefix }}%
\providecommand \urlprefix  [0]{URL }%
\providecommand \Eprint [0]{\href }%
\providecommand \doibase [0]{http://dx.doi.org/}%
\providecommand \selectlanguage [0]{\@gobble}%
\providecommand \bibinfo  [0]{\@secondoftwo}%
\providecommand \bibfield  [0]{\@secondoftwo}%
\providecommand \translation [1]{[#1]}%
\providecommand \BibitemOpen [0]{}%
\providecommand \bibitemStop [0]{}%
\providecommand \bibitemNoStop [0]{.\EOS\space}%
\providecommand \EOS [0]{\spacefactor3000\relax}%
\providecommand \BibitemShut  [1]{\csname bibitem#1\endcsname}%
\let\auto@bib@innerbib\@empty
\bibitem [{\citenamefont {Deane}\ and\ \citenamefont
  {Stokes}(2002)}]{oceanwave_bubble}%
  \BibitemOpen
  \bibfield  {author} {\bibinfo {author} {\bibfnamefont {G.~B.}\ \bibnamefont
  {Deane}}\ and\ \bibinfo {author} {\bibfnamefont {M.~D.}\ \bibnamefont
  {Stokes}},\ }\href@noop {} {\bibfield  {journal} {\bibinfo  {journal}
  {Nature}\ }\textbf {\bibinfo {volume} {418}},\ \bibinfo {pages} {839}
  (\bibinfo {year} {2002})}\BibitemShut {NoStop}%
\bibitem [{\citenamefont {Craig}\ \emph
  {et~al.}(1993{\natexlab{a}})\citenamefont {Craig}, \citenamefont {Ninham},\
  and\ \citenamefont {Pashley}}]{craig1993effect}%
  \BibitemOpen
  \bibfield  {author} {\bibinfo {author} {\bibfnamefont {V.~S.}\ \bibnamefont
  {Craig}}, \bibinfo {author} {\bibfnamefont {B.~W.}\ \bibnamefont {Ninham}}, \
  and\ \bibinfo {author} {\bibfnamefont {R.~M.}\ \bibnamefont {Pashley}},\
  }\href@noop {} {\bibfield  {journal} {\bibinfo  {journal} {J. Phys. Chem.}\
  }\textbf {\bibinfo {volume} {97}},\ \bibinfo {pages} {10192} (\bibinfo {year}
  {1993}{\natexlab{a}})}\BibitemShut {NoStop}%
\bibitem [{\citenamefont {Craig}(2011)}]{craig2011hydration}%
  \BibitemOpen
  \bibfield  {author} {\bibinfo {author} {\bibfnamefont {V.~S.}\ \bibnamefont
  {Craig}},\ }\href@noop {} {\bibfield  {journal} {\bibinfo  {journal} {Curr.
  Opin. Colloid Interface Sci.}\ }\textbf {\bibinfo {volume} {16}},\ \bibinfo
  {pages} {597} (\bibinfo {year} {2011})}\BibitemShut {NoStop}%
\bibitem [{\citenamefont {Jungwirth}\ and\ \citenamefont
  {Tobias}(2006)}]{jungwirth_specific_ion2006}%
  \BibitemOpen
  \bibfield  {author} {\bibinfo {author} {\bibfnamefont {P.}~\bibnamefont
  {Jungwirth}}\ and\ \bibinfo {author} {\bibfnamefont {D.~J.}\ \bibnamefont
  {Tobias}},\ }\href {\doibase 10.1021/cr0403741} {\bibfield  {journal}
  {\bibinfo  {journal} {Chem. Rev.}\ }\textbf {\bibinfo {volume} {106}},\
  \bibinfo {pages} {1259} (\bibinfo {year} {2006})},\ \bibinfo {note} {pMID:
  16608180},\ \Eprint {http://arxiv.org/abs/https://doi.org/10.1021/cr0403741}
  {https://doi.org/10.1021/cr0403741} \BibitemShut {NoStop}%
\bibitem [{\citenamefont {Jungwirth}\ and\ \citenamefont
  {Cremer}(2014)}]{jungwirth2014beyondhofmeister}%
  \BibitemOpen
  \bibfield  {author} {\bibinfo {author} {\bibfnamefont {P.}~\bibnamefont
  {Jungwirth}}\ and\ \bibinfo {author} {\bibfnamefont {P.~S.}\ \bibnamefont
  {Cremer}},\ }\href@noop {} {\bibfield  {journal} {\bibinfo  {journal} {Nat.
  Chem.}\ }\textbf {\bibinfo {volume} {6}},\ \bibinfo {pages} {261} (\bibinfo
  {year} {2014})}\BibitemShut {NoStop}%
\bibitem [{\citenamefont {Pegram}\ and\ \citenamefont
  {Record}(2007)}]{pegram2007hofmeister}%
  \BibitemOpen
  \bibfield  {author} {\bibinfo {author} {\bibfnamefont {L.~M.}\ \bibnamefont
  {Pegram}}\ and\ \bibinfo {author} {\bibfnamefont {M.~T.}\ \bibnamefont
  {Record}},\ }\href@noop {} {\bibfield  {journal} {\bibinfo  {journal} {J.
  Phys. Chem. B}\ }\textbf {\bibinfo {volume} {111}},\ \bibinfo {pages} {5411}
  (\bibinfo {year} {2007})}\BibitemShut {NoStop}%
\bibitem [{\citenamefont {dos Santos}\ \emph {et~al.}(2010)\citenamefont {dos
  Santos}, \citenamefont {Diehl},\ and\ \citenamefont
  {Levin}}]{levin_langmuir}%
  \BibitemOpen
  \bibfield  {author} {\bibinfo {author} {\bibfnamefont {A.~P.}\ \bibnamefont
  {dos Santos}}, \bibinfo {author} {\bibfnamefont {A.}~\bibnamefont {Diehl}}, \
  and\ \bibinfo {author} {\bibfnamefont {Y.}~\bibnamefont {Levin}},\ }\href
  {\doibase 10.1021/la100604k} {\bibfield  {journal} {\bibinfo  {journal}
  {Langmuir}\ }\textbf {\bibinfo {volume} {26}},\ \bibinfo {pages} {10778}
  (\bibinfo {year} {2010})},\ \Eprint
  {http://arxiv.org/abs/https://doi.org/10.1021/la100604k}
  {https://doi.org/10.1021/la100604k} \BibitemShut {NoStop}%
\bibitem [{\citenamefont {Levin}\ \emph
  {et~al.}(2009{\natexlab{a}})\citenamefont {Levin}, \citenamefont {dos
  Santos},\ and\ \citenamefont {Diehl}}]{levin_prl_ionsatinterface}%
  \BibitemOpen
  \bibfield  {author} {\bibinfo {author} {\bibfnamefont {Y.}~\bibnamefont
  {Levin}}, \bibinfo {author} {\bibfnamefont {A.~P.}\ \bibnamefont {dos
  Santos}}, \ and\ \bibinfo {author} {\bibfnamefont {A.}~\bibnamefont
  {Diehl}},\ }\href {\doibase 10.1103/PhysRevLett.103.257802} {\bibfield
  {journal} {\bibinfo  {journal} {Phys. Rev. Lett.}\ }\textbf {\bibinfo
  {volume} {103}},\ \bibinfo {pages} {257802} (\bibinfo {year}
  {2009}{\natexlab{a}})}\BibitemShut {NoStop}%
\bibitem [{\citenamefont {Liu}\ \emph {et~al.}(2019{\natexlab{a}})\citenamefont
  {Liu}, \citenamefont {Manica}, \citenamefont {Liu}, \citenamefont
  {Klaseboer}, \citenamefont {Xu},\ and\ \citenamefont
  {Xie}}]{liu2019coalescence}%
  \BibitemOpen
  \bibfield  {author} {\bibinfo {author} {\bibfnamefont {B.}~\bibnamefont
  {Liu}}, \bibinfo {author} {\bibfnamefont {R.}~\bibnamefont {Manica}},
  \bibinfo {author} {\bibfnamefont {Q.}~\bibnamefont {Liu}}, \bibinfo {author}
  {\bibfnamefont {E.}~\bibnamefont {Klaseboer}}, \bibinfo {author}
  {\bibfnamefont {Z.}~\bibnamefont {Xu}}, \ and\ \bibinfo {author}
  {\bibfnamefont {G.}~\bibnamefont {Xie}},\ }\href@noop {} {\bibfield
  {journal} {\bibinfo  {journal} {Phys. Rev. Lett.}\ }\textbf {\bibinfo
  {volume} {122}},\ \bibinfo {pages} {194501} (\bibinfo {year}
  {2019}{\natexlab{a}})}\BibitemShut {NoStop}%
\bibitem [{\citenamefont {Bird}\ \emph {et~al.}(2010)\citenamefont {Bird},
  \citenamefont {De~Ruiter}, \citenamefont {Courbin},\ and\ \citenamefont
  {Stone}}]{bird2010daughterbubble}%
  \BibitemOpen
  \bibfield  {author} {\bibinfo {author} {\bibfnamefont {J.~C.}\ \bibnamefont
  {Bird}}, \bibinfo {author} {\bibfnamefont {R.}~\bibnamefont {De~Ruiter}},
  \bibinfo {author} {\bibfnamefont {L.}~\bibnamefont {Courbin}}, \ and\
  \bibinfo {author} {\bibfnamefont {H.~A.}\ \bibnamefont {Stone}},\ }\href@noop
  {} {\bibfield  {journal} {\bibinfo  {journal} {Nature}\ }\textbf {\bibinfo
  {volume} {465}},\ \bibinfo {pages} {759} (\bibinfo {year}
  {2010})}\BibitemShut {NoStop}%
\bibitem [{\citenamefont {Ninham}\ \emph {et~al.}(2017)\citenamefont {Ninham},
  \citenamefont {Pashley},\ and\ \citenamefont {Nostro}}]{ninham2017surface}%
  \BibitemOpen
  \bibfield  {author} {\bibinfo {author} {\bibfnamefont {B.~W.}\ \bibnamefont
  {Ninham}}, \bibinfo {author} {\bibfnamefont {R.~M.}\ \bibnamefont {Pashley}},
  \ and\ \bibinfo {author} {\bibfnamefont {P.~L.}\ \bibnamefont {Nostro}},\
  }\href@noop {} {\bibfield  {journal} {\bibinfo  {journal} {Current Opinion in
  Colloid \& Interface Science}\ }\textbf {\bibinfo {volume} {27}},\ \bibinfo
  {pages} {25} (\bibinfo {year} {2017})}\BibitemShut {NoStop}%
\bibitem [{\citenamefont {Angulo}\ \emph {et~al.}(2020)\citenamefont {Angulo},
  \citenamefont {[van~der Linde]}, \citenamefont {Gardeniers}, \citenamefont
  {Modestino},\ and\ \citenamefont {Rivas]}}]{bubble_electrochemical}%
  \BibitemOpen
  \bibfield  {author} {\bibinfo {author} {\bibfnamefont {A.}~\bibnamefont
  {Angulo}}, \bibinfo {author} {\bibfnamefont {P.}~\bibnamefont {[van~der
  Linde]}}, \bibinfo {author} {\bibfnamefont {H.}~\bibnamefont {Gardeniers}},
  \bibinfo {author} {\bibfnamefont {M.}~\bibnamefont {Modestino}}, \ and\
  \bibinfo {author} {\bibfnamefont {D.~F.}\ \bibnamefont {Rivas]}},\ }\href
  {\doibase https://doi.org/10.1016/j.joule.2020.01.005} {\bibfield  {journal}
  {\bibinfo  {journal} {Joule}\ }\textbf {\bibinfo {volume} {4}},\ \bibinfo
  {pages} {555 } (\bibinfo {year} {2020})}\BibitemShut {NoStop}%
\bibitem [{\citenamefont {Yoon}(1993)}]{yoon1993microbubble}%
  \BibitemOpen
  \bibfield  {author} {\bibinfo {author} {\bibfnamefont {R.-H.}\ \bibnamefont
  {Yoon}},\ }\href@noop {} {\bibfield  {journal} {\bibinfo  {journal} {Miner.
  Eng.}\ }\textbf {\bibinfo {volume} {6}},\ \bibinfo {pages} {619} (\bibinfo
  {year} {1993})}\BibitemShut {NoStop}%
\bibitem [{\citenamefont {Craig}\ \emph
  {et~al.}(1993{\natexlab{b}})\citenamefont {Craig}, \citenamefont {Ninham},\
  and\ \citenamefont {Pashley}}]{Craig1993}%
  \BibitemOpen
  \bibfield  {author} {\bibinfo {author} {\bibfnamefont {V.~S.}\ \bibnamefont
  {Craig}}, \bibinfo {author} {\bibfnamefont {B.~W.}\ \bibnamefont {Ninham}}, \
  and\ \bibinfo {author} {\bibfnamefont {R.~M.}\ \bibnamefont {Pashley}},\
  }\href@noop {} {\bibfield  {journal} {\bibinfo  {journal} {The Journal of
  Physical Chemistry}\ }\textbf {\bibinfo {volume} {97}},\ \bibinfo {pages}
  {10192} (\bibinfo {year} {1993}{\natexlab{b}})}\BibitemShut {NoStop}%
\bibitem [{\citenamefont {Marrucci}\ and\ \citenamefont
  {Nicodemo}(1967)}]{marrucci1967coalescence}%
  \BibitemOpen
  \bibfield  {author} {\bibinfo {author} {\bibfnamefont {G.}~\bibnamefont
  {Marrucci}}\ and\ \bibinfo {author} {\bibfnamefont {L.}~\bibnamefont
  {Nicodemo}},\ }\href@noop {} {\bibfield  {journal} {\bibinfo  {journal}
  {Chem. Eng. Sci.}\ }\textbf {\bibinfo {volume} {22}},\ \bibinfo {pages}
  {1257} (\bibinfo {year} {1967})}\BibitemShut {NoStop}%
\bibitem [{\citenamefont {Christenson}\ \emph {et~al.}(2008)\citenamefont
  {Christenson}, \citenamefont {Bowen}, \citenamefont {Carlton}, \citenamefont
  {Denne},\ and\ \citenamefont {Lu}}]{christenson2008electrolytes}%
  \BibitemOpen
  \bibfield  {author} {\bibinfo {author} {\bibfnamefont {H.}~\bibnamefont
  {Christenson}}, \bibinfo {author} {\bibfnamefont {R.}~\bibnamefont {Bowen}},
  \bibinfo {author} {\bibfnamefont {J.}~\bibnamefont {Carlton}}, \bibinfo
  {author} {\bibfnamefont {J.}~\bibnamefont {Denne}}, \ and\ \bibinfo {author}
  {\bibfnamefont {Y.}~\bibnamefont {Lu}},\ }\href@noop {} {\bibfield  {journal}
  {\bibinfo  {journal} {J. Phys. Chem. C}\ }\textbf {\bibinfo {volume} {112}},\
  \bibinfo {pages} {794} (\bibinfo {year} {2008})}\BibitemShut {NoStop}%
\bibitem [{\citenamefont {Kirkpatrick}\ and\ \citenamefont
  {Lockett}(1974)}]{kirkpatrick1974influence}%
  \BibitemOpen
  \bibfield  {author} {\bibinfo {author} {\bibfnamefont {R.}~\bibnamefont
  {Kirkpatrick}}\ and\ \bibinfo {author} {\bibfnamefont {M.}~\bibnamefont
  {Lockett}},\ }\href@noop {} {\bibfield  {journal} {\bibinfo  {journal} {Chem.
  Eng. Sci.}\ }\textbf {\bibinfo {volume} {29}},\ \bibinfo {pages} {2363}
  (\bibinfo {year} {1974})}\BibitemShut {NoStop}%
\bibitem [{\citenamefont {Liu}\ \emph {et~al.}(2019{\natexlab{b}})\citenamefont
  {Liu}, \citenamefont {Manica}, \citenamefont {Liu}, \citenamefont
  {Klaseboer},\ and\ \citenamefont {Xu}}]{liu2019coalescencejpcl}%
  \BibitemOpen
  \bibfield  {author} {\bibinfo {author} {\bibfnamefont {B.}~\bibnamefont
  {Liu}}, \bibinfo {author} {\bibfnamefont {R.}~\bibnamefont {Manica}},
  \bibinfo {author} {\bibfnamefont {Q.}~\bibnamefont {Liu}}, \bibinfo {author}
  {\bibfnamefont {E.}~\bibnamefont {Klaseboer}}, \ and\ \bibinfo {author}
  {\bibfnamefont {Z.}~\bibnamefont {Xu}},\ }\href@noop {} {\bibfield  {journal}
  {\bibinfo  {journal} {J. Phys. Chem. Lett.}\ }\textbf {\bibinfo {volume}
  {10}},\ \bibinfo {pages} {5662} (\bibinfo {year}
  {2019}{\natexlab{b}})}\BibitemShut {NoStop}%
\bibitem [{\citenamefont {Pegram}\ and\ \citenamefont
  {Record}(2006)}]{pegram2006partitioning}%
  \BibitemOpen
  \bibfield  {author} {\bibinfo {author} {\bibfnamefont {L.~M.}\ \bibnamefont
  {Pegram}}\ and\ \bibinfo {author} {\bibfnamefont {M.~T.}\ \bibnamefont
  {Record}},\ }\href@noop {} {\bibfield  {journal} {\bibinfo  {journal} {Proc.
  Natl. Acad. Sci. U.S.A.}\ }\textbf {\bibinfo {volume} {103}},\ \bibinfo
  {pages} {14278} (\bibinfo {year} {2006})}\BibitemShut {NoStop}%
\bibitem [{\citenamefont {Weissenborn}\ and\ \citenamefont
  {Pugh}(1996)}]{ionhydration_coalescence_cacl2_gascontent}%
  \BibitemOpen
  \bibfield  {author} {\bibinfo {author} {\bibfnamefont {P.~K.}\ \bibnamefont
  {Weissenborn}}\ and\ \bibinfo {author} {\bibfnamefont {R.~J.}\ \bibnamefont
  {Pugh}},\ }\href@noop {} {\bibfield  {journal} {\bibinfo  {journal} {J.
  Colloid Interface Sci.}\ }\textbf {\bibinfo {volume} {184}},\ \bibinfo
  {pages} {550} (\bibinfo {year} {1996})}\BibitemShut {NoStop}%
\bibitem [{\citenamefont {Levin}\ \emph
  {et~al.}(2009{\natexlab{b}})\citenamefont {Levin}, \citenamefont
  {Dos~Santos},\ and\ \citenamefont {Diehl}}]{levin2009ions}%
  \BibitemOpen
  \bibfield  {author} {\bibinfo {author} {\bibfnamefont {Y.}~\bibnamefont
  {Levin}}, \bibinfo {author} {\bibfnamefont {A.~P.}\ \bibnamefont
  {Dos~Santos}}, \ and\ \bibinfo {author} {\bibfnamefont {A.}~\bibnamefont
  {Diehl}},\ }\href@noop {} {\bibfield  {journal} {\bibinfo  {journal} {Phys.
  Rev. Lett.}\ }\textbf {\bibinfo {volume} {103}},\ \bibinfo {pages} {257802}
  (\bibinfo {year} {2009}{\natexlab{b}})}\BibitemShut {NoStop}%
\bibitem [{\citenamefont {Chan}\ \emph {et~al.}(2011)\citenamefont {Chan},
  \citenamefont {Klaseboer},\ and\ \citenamefont {Manica}}]{Chan2011a}%
  \BibitemOpen
  \bibfield  {author} {\bibinfo {author} {\bibfnamefont {D.~Y.~C.}\
  \bibnamefont {Chan}}, \bibinfo {author} {\bibfnamefont {E.}~\bibnamefont
  {Klaseboer}}, \ and\ \bibinfo {author} {\bibfnamefont {R.}~\bibnamefont
  {Manica}},\ }\href {\doibase 10.1039/C0SM00812E} {\bibfield  {journal}
  {\bibinfo  {journal} {Soft Matter}\ }\textbf {\bibinfo {volume} {7}},\
  \bibinfo {pages} {2235} (\bibinfo {year} {2011})}\BibitemShut {NoStop}%
\bibitem [{\citenamefont {Davis}\ \emph {et~al.}(1989)\citenamefont {Davis},
  \citenamefont {Schonberg},\ and\ \citenamefont
  {Rallison}}]{davis1989lubrication}%
  \BibitemOpen
  \bibfield  {author} {\bibinfo {author} {\bibfnamefont {R.~H.}\ \bibnamefont
  {Davis}}, \bibinfo {author} {\bibfnamefont {J.~A.}\ \bibnamefont
  {Schonberg}}, \ and\ \bibinfo {author} {\bibfnamefont {J.~M.}\ \bibnamefont
  {Rallison}},\ }\href {\doibase 10.1063/1.857525} {\bibfield  {journal}
  {\bibinfo  {journal} {Physics of Fluids A: Fluid Dynamics}\ }\textbf
  {\bibinfo {volume} {1}},\ \bibinfo {pages} {77} (\bibinfo {year}
  {1989})}\BibitemShut {NoStop}%
\bibitem [{\citenamefont {Henry}\ \emph {et~al.}(2007)\citenamefont {Henry},
  \citenamefont {Dalton}, \citenamefont {Scruton},\ and\ \citenamefont
  {Craig}}]{henry2007ion}%
  \BibitemOpen
  \bibfield  {author} {\bibinfo {author} {\bibfnamefont {C.~L.}\ \bibnamefont
  {Henry}}, \bibinfo {author} {\bibfnamefont {C.~N.}\ \bibnamefont {Dalton}},
  \bibinfo {author} {\bibfnamefont {L.}~\bibnamefont {Scruton}}, \ and\
  \bibinfo {author} {\bibfnamefont {V.~S.}\ \bibnamefont {Craig}},\ }\href@noop
  {} {\bibfield  {journal} {\bibinfo  {journal} {J. Phys. Chem. C}\ }\textbf
  {\bibinfo {volume} {111}},\ \bibinfo {pages} {1015} (\bibinfo {year}
  {2007})}\BibitemShut {NoStop}%
\bibitem [{\citenamefont {Venkateshwaran}\ \emph {et~al.}(2014)\citenamefont
  {Venkateshwaran}, \citenamefont {Vembanur},\ and\ \citenamefont
  {Garde}}]{PNAS_ion_ion_correlation_interface}%
  \BibitemOpen
  \bibfield  {author} {\bibinfo {author} {\bibfnamefont {V.}~\bibnamefont
  {Venkateshwaran}}, \bibinfo {author} {\bibfnamefont {S.}~\bibnamefont
  {Vembanur}}, \ and\ \bibinfo {author} {\bibfnamefont {S.}~\bibnamefont
  {Garde}},\ }\href@noop {} {\bibfield  {journal} {\bibinfo  {journal} {Proc.
  Natl. Acad. Sci. U.S.A.}\ }\textbf {\bibinfo {volume} {111}},\ \bibinfo
  {pages} {8729} (\bibinfo {year} {2014})}\BibitemShut {NoStop}%
\bibitem [{\citenamefont {Henry}\ and\ \citenamefont
  {Craig}(2008)}]{henry_non_aqueous}%
  \BibitemOpen
  \bibfield  {author} {\bibinfo {author} {\bibfnamefont {C.~L.}\ \bibnamefont
  {Henry}}\ and\ \bibinfo {author} {\bibfnamefont {V.~S.}\ \bibnamefont
  {Craig}},\ }\href@noop {} {\bibfield  {journal} {\bibinfo  {journal}
  {Langmuir}\ }\textbf {\bibinfo {volume} {24}},\ \bibinfo {pages} {7979}
  (\bibinfo {year} {2008})}\BibitemShut {NoStop}%
\bibitem [{\citenamefont {Bruce}\ and\ \citenamefont {van~der
  Vegt}(2019)}]{jacs_salt_noadditive}%
  \BibitemOpen
  \bibfield  {author} {\bibinfo {author} {\bibfnamefont {E.~E.}\ \bibnamefont
  {Bruce}}\ and\ \bibinfo {author} {\bibfnamefont {N.~F.}\ \bibnamefont
  {van~der Vegt}},\ }\href@noop {} {\bibfield  {journal} {\bibinfo  {journal}
  {J. Am. Chem. Soc.}\ }\textbf {\bibinfo {volume} {141}},\ \bibinfo {pages}
  {12948} (\bibinfo {year} {2019})}\BibitemShut {NoStop}%
\bibitem [{\citenamefont {Lis}\ \emph {et~al.}(2014)\citenamefont {Lis},
  \citenamefont {Backus}, \citenamefont {Hunger}, \citenamefont {Parekh},\ and\
  \citenamefont {Bonn}}]{liquidflow_solid_science}%
  \BibitemOpen
  \bibfield  {author} {\bibinfo {author} {\bibfnamefont {D.}~\bibnamefont
  {Lis}}, \bibinfo {author} {\bibfnamefont {E.~H.}\ \bibnamefont {Backus}},
  \bibinfo {author} {\bibfnamefont {J.}~\bibnamefont {Hunger}}, \bibinfo
  {author} {\bibfnamefont {S.~H.}\ \bibnamefont {Parekh}}, \ and\ \bibinfo
  {author} {\bibfnamefont {M.}~\bibnamefont {Bonn}},\ }\href@noop {} {\bibfield
   {journal} {\bibinfo  {journal} {Science}\ }\textbf {\bibinfo {volume}
  {344}},\ \bibinfo {pages} {1138} (\bibinfo {year} {2014})}\BibitemShut
  {NoStop}%
\bibitem [{\citenamefont {Vakarelski}\ \emph {et~al.}(2010)\citenamefont
  {Vakarelski}, \citenamefont {Manica}, \citenamefont {Tang}, \citenamefont
  {O’Shea}, \citenamefont {Stevens}, \citenamefont {Grieser}, \citenamefont
  {Dagastine},\ and\ \citenamefont {Chan}}]{vakarelski2010dynamic}%
  \BibitemOpen
  \bibfield  {author} {\bibinfo {author} {\bibfnamefont {I.~U.}\ \bibnamefont
  {Vakarelski}}, \bibinfo {author} {\bibfnamefont {R.}~\bibnamefont {Manica}},
  \bibinfo {author} {\bibfnamefont {X.}~\bibnamefont {Tang}}, \bibinfo {author}
  {\bibfnamefont {S.~J.}\ \bibnamefont {O’Shea}}, \bibinfo {author}
  {\bibfnamefont {G.~W.}\ \bibnamefont {Stevens}}, \bibinfo {author}
  {\bibfnamefont {F.}~\bibnamefont {Grieser}}, \bibinfo {author} {\bibfnamefont
  {R.~R.}\ \bibnamefont {Dagastine}}, \ and\ \bibinfo {author} {\bibfnamefont
  {D.~Y.}\ \bibnamefont {Chan}},\ }\href@noop {} {\bibfield  {journal}
  {\bibinfo  {journal} {Proc. Natl. Acad. Sci. U.S.A.}\ }\textbf {\bibinfo
  {volume} {107}},\ \bibinfo {pages} {11177} (\bibinfo {year}
  {2010})}\BibitemShut {NoStop}%
\end{thebibliography}%


\begin{thebibliography}{28}%
	\makeatletter
	\providecommand \@ifxundefined [1]{%
		\@ifx{#1\undefined}
	}%
	\providecommand \@ifnum [1]{%
		\ifnum #1\expandafter \@firstoftwo
		\else \expandafter \@secondoftwo
		\fi
	}%
	\providecommand \@ifx [1]{%
		\ifx #1\expandafter \@firstoftwo
		\else \expandafter \@secondoftwo
		\fi
	}%
	\providecommand \natexlab [1]{#1}%
	\providecommand \enquote  [1]{``#1''}%
	\providecommand \bibnamefont  [1]{#1}%
	\providecommand \bibfnamefont [1]{#1}%
	\providecommand \citenamefont [1]{#1}%
	\providecommand \href@noop [0]{\@secondoftwo}%
	\providecommand \href [0]{\begingroup \@sanitize@url \@href}%
	\providecommand \@href[1]{\@@startlink{#1}\@@href}%
	\providecommand \@@href[1]{\endgroup#1\@@endlink}%
	\providecommand \@sanitize@url [0]{\catcode `\\12\catcode `\$12\catcode
		`\&12\catcode `\#12\catcode `\^12\catcode `\_12\catcode `\%12\relax}%
	\providecommand \@@startlink[1]{}%
	\providecommand \@@endlink[0]{}%
	\providecommand \url  [0]{\begingroup\@sanitize@url \@url }%
	\providecommand \@url [1]{\endgroup\@href {#1}{\urlprefix }}%
	\providecommand \urlprefix  [0]{URL }%
	\providecommand \Eprint [0]{\href }%
	\providecommand \doibase [0]{http://dx.doi.org/}%
	\providecommand \selectlanguage [0]{\@gobble}%
	\providecommand \bibinfo  [0]{\@secondoftwo}%
	\providecommand \bibfield  [0]{\@secondoftwo}%
	\providecommand \translation [1]{[#1]}%
	\providecommand \BibitemOpen [0]{}%
	\providecommand \bibitemStop [0]{}%
	\providecommand \bibitemNoStop [0]{.\EOS\space}%
	\providecommand \EOS [0]{\spacefactor3000\relax}%
	\providecommand \BibitemShut  [1]{\csname bibitem#1\endcsname}%
	\let\auto@bib@innerbib\@empty
	\bibitem [{\citenamefont {Verschoof}\ \emph {et~al.}(2016)\citenamefont
		{Verschoof}, \citenamefont {{van der Veen}}, \citenamefont {Sun},\ and\
		\citenamefont {Lohse}}]{Verschoof2016}%
	\BibitemOpen
	\bibfield  {author} {\bibinfo {author} {\bibfnamefont {R.~A.}\ \bibnamefont
			{Verschoof}}, \bibinfo {author} {\bibfnamefont {R.~C.~A.}\ \bibnamefont {{van
					der Veen}}}, \bibinfo {author} {\bibfnamefont {C.}~\bibnamefont {Sun}}, \
		and\ \bibinfo {author} {\bibfnamefont {D.}~\bibnamefont {Lohse}},\ }\href
	{\doibase 10.1103/PhysRevLett.117.104502} {\bibfield  {journal} {\bibinfo
			{journal} {Phys. Rev. Lett.}\ }\textbf {\bibinfo {volume} {117}},\ \bibinfo
		{pages} {104502} (\bibinfo {year} {2016})}\  \BibitemShut {NoStop}%
	\bibitem [{\citenamefont {Yoon}(1993)}]{yoon1993microbubble}%
	\BibitemOpen
	\bibfield  {author} {\bibinfo {author} {\bibfnamefont {R.~H.}\ \bibnamefont
			{Yoon}},\ }\href@noop {} {\bibfield  {journal} {\bibinfo  {journal} {Miner.
				Eng.}\ }\textbf {\bibinfo {volume} {6}},\ \bibinfo {pages} {619} (\bibinfo
		{year} {1993})}\BibitemShut {NoStop}%
	\bibitem [{\citenamefont {Horn}\ \emph {et~al.}(2011)\citenamefont {Horn},
		\citenamefont {{Del Castillo}},\ and\ \citenamefont {Ohnishi}}]{Horn2011a}%
	\BibitemOpen
	\bibfield  {author} {\bibinfo {author} {\bibfnamefont {R.~G.}\ \bibnamefont
			{Horn}}, \bibinfo {author} {\bibfnamefont {L.~A.}\ \bibnamefont {{Del
					Castillo}}}, \ and\ \bibinfo {author} {\bibfnamefont {S.}~\bibnamefont
			{Ohnishi}},\ }\href {\doibase 10.1016/j.cis.2011.05.006} {\bibfield
		{journal} {\bibinfo  {journal} {Adv. Colloid Interface Sci.}\ }\textbf
		{\bibinfo {volume} {168}},\ \bibinfo {pages} {85} (\bibinfo {year}
		{2011})}\BibitemShut {NoStop}%
	\bibitem [{\citenamefont {Liu}\ \emph {et~al.}(2018)\citenamefont {Liu},
		\citenamefont {Manica}, \citenamefont {Zhang}, \citenamefont {Bussonnière},
		\citenamefont {Xu}, \citenamefont {Xie},\ and\ \citenamefont
		{Liu}}]{liu2018dynamic}%
	\BibitemOpen
	\bibfield  {author} {\bibinfo {author} {\bibfnamefont {B.}~\bibnamefont
			{Liu}}, \bibinfo {author} {\bibfnamefont {R.}~\bibnamefont {Manica}},
		\bibinfo {author} {\bibfnamefont {X.}~\bibnamefont {Zhang}}, \bibinfo
		{author} {\bibfnamefont {A.}~\bibnamefont {Bussonni\`ere}}, \bibinfo {author}
		{\bibfnamefont {Z.}~\bibnamefont {Xu}}, \bibinfo {author} {\bibfnamefont
			{G.}~\bibnamefont {Xie}}, \ and\ \bibinfo {author} {\bibfnamefont
			{Q.}~\bibnamefont {Liu}},\ }\href@noop {} {\bibfield  {journal} {\bibinfo
			{journal} {Langmuir}\ }\textbf {\bibinfo {volume} {34}},\ \bibinfo {pages}
		{11667} (\bibinfo {year} {2018})}\BibitemShut {NoStop}%
	\bibitem [{\citenamefont {Kirkpatrick}\ and\ \citenamefont
		{Lockett}(1974)}]{kirkpatrick1974influence}%
	\BibitemOpen
	\bibfield  {author} {\bibinfo {author} {\bibfnamefont {R.}~\bibnamefont
			{Kirkpatrick}}\ and\ \bibinfo {author} {\bibfnamefont {M.}~\bibnamefont
			{Lockett}},\ }\href@noop {} {\bibfield  {journal} {\bibinfo  {journal} {Chem.
				Eng. Sci.}\ }\textbf {\bibinfo {volume} {29}},\ \bibinfo {pages} {2363}
		(\bibinfo {year} {1974})}\BibitemShut {NoStop}%
	\bibitem [{\citenamefont {Langevin}(2015)}]{Langevin2015e}%
	\BibitemOpen
	\bibfield  {author} {\bibinfo {author} {\bibfnamefont {D.}~\bibnamefont
			{Langevin}},\ }\href {\doibase 10.1016/j.cocis.2015.03.005} {\bibfield
		{journal} {\bibinfo  {journal} {Curr. Opin. Colloid Interface Sci.}\ }\textbf
		{\bibinfo {volume} {20}},\ \bibinfo {pages} {92} (\bibinfo {year}
		{2015})}\BibitemShut {NoStop}%
	\bibitem [{\citenamefont {Yaminsky}\ \emph
		{et~al.}(2010{\natexlab{a}})\citenamefont {Yaminsky}, \citenamefont
		{Ohnishi}, \citenamefont {Vogler},\ and\ \citenamefont
		{Horn}}]{yaminsky2010stability}%
	\BibitemOpen
	\bibfield  {author} {\bibinfo {author} {\bibfnamefont {V.~V.}\ \bibnamefont
			{Yaminsky}}, \bibinfo {author} {\bibfnamefont {S.}~\bibnamefont {Ohnishi}},
		\bibinfo {author} {\bibfnamefont {E.~A.}\ \bibnamefont {Vogler}}, \ and\
		\bibinfo {author} {\bibfnamefont {R.~G.}\ \bibnamefont {Horn}},\ }\href@noop
	{} {\bibfield  {journal} {\bibinfo  {journal} {Langmuir}\ }\textbf {\bibinfo
			{volume} {26}},\ \bibinfo {pages} {8061} (\bibinfo {year}
		{2010}{\natexlab{a}})}\BibitemShut {NoStop}%
\bibitem [{\citenamefont {Vakarelski}\ \emph {et~al.}(2018)\citenamefont
	{Vakarelski}, \citenamefont {Manica}, \citenamefont {Li}, \citenamefont
	{Basheva}, \citenamefont {Chan},\ and\ \citenamefont
	{Thoroddsen}}]{Vakarelski2018}%
\BibitemOpen
\bibfield  {author} {\bibinfo {author} {\bibfnamefont {I.~U.}\ \bibnamefont
		{Vakarelski}}, \bibinfo {author} {\bibfnamefont {R.}~\bibnamefont {Manica}},
	\bibinfo {author} {\bibfnamefont {E.~Q.}\ \bibnamefont {Li}}, \bibinfo
	{author} {\bibfnamefont {E.~S.}\ \bibnamefont {Basheva}}, \bibinfo {author}
	{\bibfnamefont {D.~Y.~C.}\ \bibnamefont {Chan}}, \ and\ \bibinfo {author}
	{\bibfnamefont {S.~T.}\ \bibnamefont {Thoroddsen}},\ }\href {\doibase
	10.1021/acs.langmuir.7b04106} {\bibfield  {journal} {\bibinfo  {journal}
		{Langmuir}\ }\textbf {\bibinfo {volume} {34}},\ \bibinfo {pages} {2096}
	(\bibinfo {year} {2018})}, \BibitemShut {NoStop}%
\bibitem [{\citenamefont {Vakarelski}\ \emph {et~al.}(2010)\citenamefont
	{Vakarelski}, \citenamefont {Manica}, \citenamefont {Tang}, \citenamefont
	{O'Shea}, \citenamefont {Stevens}, \citenamefont {Grieser}, \citenamefont
	{Dagastine},\ and\ \citenamefont {Chan}}]{vakarelski2010}%
\BibitemOpen
\bibfield  {author} {\bibinfo {author} {\bibfnamefont {I.~U.}\ \bibnamefont
		{Vakarelski}}, \bibinfo {author} {\bibfnamefont {R.}~\bibnamefont {Manica}},
	\bibinfo {author} {\bibfnamefont {X.}~\bibnamefont {Tang}}, \bibinfo {author}
	{\bibfnamefont {S.~J.}\ \bibnamefont {O'Shea}}, \bibinfo {author}
	{\bibfnamefont {G.~W.}\ \bibnamefont {Stevens}}, \bibinfo {author}
	{\bibfnamefont {F.}~\bibnamefont {Grieser}}, \bibinfo {author} {\bibfnamefont
		{R.~R.}\ \bibnamefont {Dagastine}}, \ and\ \bibinfo {author} {\bibfnamefont
		{D.~Y.}\ \bibnamefont {Chan}},\ }\href@noop {} {\bibfield  {journal}
	{\bibinfo  {journal} {Proc. Natl. Acad. Sci. U. S. A.}\ }\textbf {\bibinfo
		{volume} {107}},\ \bibinfo {pages} {11177} (\bibinfo {year}
	{2010})}\BibitemShut {NoStop}%
	\bibitem [{\citenamefont {Sch{\"a}ffel}\ \emph {et~al.}(2016)\citenamefont
		{Sch{\"a}ffel}, \citenamefont {Koynov}, \citenamefont {Vollmer},
		\citenamefont {Butt},\ and\ \citenamefont
		{Sch{\"o}necker}}]{schaffel2016local}%
	\BibitemOpen
	\bibfield  {author} {\bibinfo {author} {\bibfnamefont {D.}~\bibnamefont
			{Sch{\"a}ffel}}, \bibinfo {author} {\bibfnamefont {K.}~\bibnamefont
			{Koynov}}, \bibinfo {author} {\bibfnamefont {D.}~\bibnamefont {Vollmer}},
		\bibinfo {author} {\bibfnamefont {H.-J.}\ \bibnamefont {Butt}}, \ and\
		\bibinfo {author} {\bibfnamefont {C.}~\bibnamefont {Sch{\"o}necker}},\
	}\href@noop {} {\bibfield  {journal} {\bibinfo  {journal} {Phys. Rev. Lett.}\
		}\textbf {\bibinfo {volume} {116}},\ \bibinfo {pages} {134501} (\bibinfo
		{year} {2016})}\BibitemShut {NoStop}%
	\bibitem [{\citenamefont {Tsai}\ \emph {et~al.}(2009)\citenamefont {Tsai},
		\citenamefont {Peters}, \citenamefont {Pirat}, \citenamefont {Wessling},
		\citenamefont {Lammertink},\ and\ \citenamefont
		{Lohse}}]{tsai2009quantifying}%
	\BibitemOpen
	\bibfield  {author} {\bibinfo {author} {\bibfnamefont {P.}~\bibnamefont
			{Tsai}}, \bibinfo {author} {\bibfnamefont {A.~M.}\ \bibnamefont {Peters}},
		\bibinfo {author} {\bibfnamefont {C.}~\bibnamefont {Pirat}}, \bibinfo
		{author} {\bibfnamefont {M.}~\bibnamefont {Wessling}}, \bibinfo {author}
		{\bibfnamefont {R.~G.}\ \bibnamefont {Lammertink}}, \ and\ \bibinfo {author}
		{\bibfnamefont {D.}~\bibnamefont {Lohse}},\ }\href@noop {} {\bibfield
		{journal} {\bibinfo  {journal} {Phys. Fluids}\ }\textbf {\bibinfo {volume}
			{21}},\ \bibinfo {pages} {112002} (\bibinfo {year} {2009})}\BibitemShut
	{NoStop}%
	\bibitem [{\citenamefont {Vakarelski}\ \emph {et~al.}(2017)\citenamefont
		{Vakarelski}, \citenamefont {Klaseboer}, \citenamefont {Jetly}, \citenamefont
		{Mansoor}, \citenamefont {Aguirre-Pablo}, \citenamefont {Chan},\ and\
		\citenamefont {Thoroddsen}}]{vakarelski2017self}%
	\BibitemOpen
	\bibfield  {author} {\bibinfo {author} {\bibfnamefont {I.~U.}\ \bibnamefont
			{Vakarelski}}, \bibinfo {author} {\bibfnamefont {E.}~\bibnamefont
			{Klaseboer}}, \bibinfo {author} {\bibfnamefont {A.}~\bibnamefont {Jetly}},
		\bibinfo {author} {\bibfnamefont {M.~M.}\ \bibnamefont {Mansoor}}, \bibinfo
		{author} {\bibfnamefont {A.~A.}\ \bibnamefont {Aguirre-Pablo}}, \bibinfo
		{author} {\bibfnamefont {D.~Y.~C.}\ \bibnamefont {Chan}}, \ and\ \bibinfo
		{author} {\bibfnamefont {S.~T.}\ \bibnamefont {Thoroddsen}},\ }\href@noop {}
	{\bibfield  {journal} {\bibinfo  {journal} {Sci. Adv.}\ }\textbf {\bibinfo
			{volume} {3}},\ \bibinfo {pages} {e1701558} (\bibinfo {year}
		{2017})}\BibitemShut {NoStop}%
	\bibitem [{\citenamefont {Chesters}\ and\ \citenamefont
		{Hofman}(1982)}]{chesters1982bubble}%
	\BibitemOpen
	\bibfield  {author} {\bibinfo {author} {\bibfnamefont {A.~K.}\ \bibnamefont
			{Chesters}}\ and\ \bibinfo {author} {\bibfnamefont {G.}~\bibnamefont
			{Hofman}},\ }in\ \href@noop {} {\emph {\bibinfo {booktitle} {Mechanics and
				Physics of Bubbles in Liquids}}}\ (\bibinfo  {publisher} {Springer},\
	\bibinfo {year} {1982})\ pp.\ \bibinfo {pages} {353--361}\BibitemShut
	{NoStop}%
	\bibitem [{\citenamefont {Del~Castillo}\ \emph {et~al.}(2011)\citenamefont
		{Del~Castillo}, \citenamefont {Ohnishi},\ and\ \citenamefont
		{Horn}}]{del2011inhibition}%
	\BibitemOpen
	\bibfield  {author} {\bibinfo {author} {\bibfnamefont {L.~A.}\ \bibnamefont
			{Del~Castillo}}, \bibinfo {author} {\bibfnamefont {S.}~\bibnamefont
			{Ohnishi}}, \ and\ \bibinfo {author} {\bibfnamefont {R.~G.}\ \bibnamefont
			{Horn}},\ }\href@noop {} {\bibfield  {journal} {\bibinfo  {journal} {J.
				Colloid Interface Sci.}\ }\textbf {\bibinfo {volume} {356}},\ \bibinfo
		{pages} {316} (\bibinfo {year} {2011})}\BibitemShut {NoStop}%
	\bibitem [{\citenamefont {Hendrix}\ \emph {et~al.}(2012)\citenamefont
		{Hendrix}, \citenamefont {Manica}, \citenamefont {Klaseboer}, \citenamefont
		{Chan},\ and\ \citenamefont {Ohl}}]{Hendrix2012}%
	\BibitemOpen
	\bibfield  {author} {\bibinfo {author} {\bibfnamefont {M.~H.~W.}\ \bibnamefont
			{Hendrix}}, \bibinfo {author} {\bibfnamefont {R.}~\bibnamefont {Manica}},
		\bibinfo {author} {\bibfnamefont {E.}~\bibnamefont {Klaseboer}}, \bibinfo
		{author} {\bibfnamefont {D.~Y.~C.}\ \bibnamefont {Chan}}, \ and\ \bibinfo
		{author} {\bibfnamefont {C.~D.}\ \bibnamefont {Ohl}},\ }\href {\doibase
		10.1103/PhysRevLett.108.247803} {\bibfield  {journal} {\bibinfo  {journal}
			{Phys. Rev. Lett.}\ }\textbf {\bibinfo {volume} {108}},\ \bibinfo {pages} {247803}
		(\bibinfo {year} {2012})}\BibitemShut {NoStop}%
	\bibitem [{\citenamefont {Parkinson}\ and\ \citenamefont
		{Ralston}(2010)}]{Parkinson2010b}%
	\BibitemOpen
	\bibfield  {author} {\bibinfo {author} {\bibfnamefont {L.}~\bibnamefont
			{Parkinson}}\ and\ \bibinfo {author} {\bibfnamefont {J.}~\bibnamefont
			{Ralston}},\ }\href {\doibase 10.1021/jp9099754} {\bibfield  {journal}
		{\bibinfo  {journal} {Journal of Physical Chemistry C}\ }\textbf {\bibinfo
			{volume} {114}},\ \bibinfo {pages} {2273} (\bibinfo {year}
		{2010})}\BibitemShut {NoStop}%
	\bibitem [{\citenamefont {Chan}\ \emph {et~al.}(2011)\citenamefont {Chan},
		\citenamefont {Klaseboer},\ and\ \citenamefont {Manica}}]{Chan2011a}%
	\BibitemOpen
	\bibfield  {author} {\bibinfo {author} {\bibfnamefont {D.~Y.~C.}\
			\bibnamefont {Chan}}, \bibinfo {author} {\bibfnamefont {E.}~\bibnamefont
			{Klaseboer}}, \ and\ \bibinfo {author} {\bibfnamefont {R.}~\bibnamefont
			{Manica}},\ }\href {\doibase 10.1016/j.cis.2010.12.001} {\bibfield  {journal}
		{\bibinfo  {journal} {Adv. Colloid Interface Sci.}\ }\textbf {\bibinfo
			{volume} {165}},\ \bibinfo {pages} {70} (\bibinfo {year} {2011})}\BibitemShut
	{NoStop}%
	\bibitem [{\citenamefont {Zhang}\ \emph {et~al.}(2016)\citenamefont {Zhang},
		\citenamefont {Tchoukov}, \citenamefont {Manica}, \citenamefont {Wang},
		\citenamefont {Liu},\ and\ \citenamefont {Xu}}]{Zhang2017b}%
	\BibitemOpen
	\bibfield  {author} {\bibinfo {author} {\bibfnamefont {X.}~\bibnamefont
			{Zhang}}, \bibinfo {author} {\bibfnamefont {P.}~\bibnamefont {Tchoukov}},
		\bibinfo {author} {\bibfnamefont {R.}~\bibnamefont {Manica}}, \bibinfo
		{author} {\bibfnamefont {L.}~\bibnamefont {Wang}}, \bibinfo {author}
		{\bibfnamefont {Q.}~\bibnamefont {Liu}}, \ and\ \bibinfo {author}
		{\bibfnamefont {Z.}~\bibnamefont {Xu}},\ }\href@noop {} {\bibfield  {journal}
		{\bibinfo  {journal} {Soft Matter}\ }\textbf {\bibinfo {volume} {12}},\
		\bibinfo {pages} {9105} (\bibinfo {year} {2016})}\BibitemShut {NoStop}%
	\bibitem [{\citenamefont {Gao}\ and\ \citenamefont
		{Pan}(2018)}]{gao2018measurement}%
	\BibitemOpen
	\bibfield  {author} {\bibinfo {author} {\bibfnamefont {Y.}~\bibnamefont
			{Gao}}\ and\ \bibinfo {author} {\bibfnamefont {L.}~\bibnamefont {Pan}},\
	}\href@noop {} {\bibfield  {journal} {\bibinfo  {journal} {Langmuir}\
		}\textbf {\bibinfo {volume} {34}},\ \bibinfo {pages} {14215} (\bibinfo {year}
		{2018})}\BibitemShut {NoStop}%
	\bibitem [{\citenamefont {Davis}\ \emph {et~al.}(1989)\citenamefont {Davis},
		\citenamefont {Schonberg},\ and\ \citenamefont
		{Rallison}}]{davis1989lubrication}%
	\BibitemOpen
	\bibfield  {author} {\bibinfo {author} {\bibfnamefont {R.~H.}\ \bibnamefont
			{Davis}}, \bibinfo {author} {\bibfnamefont {J.~A.}\ \bibnamefont
			{Schonberg}}, \ and\ \bibinfo {author} {\bibfnamefont {J.~M.}\ \bibnamefont
			{Rallison}},\ }\href@noop {} {\bibfield  {journal} {\bibinfo  {journal}
			{Phys. Fluids A: Fluid Dynamics}\ }\textbf {\bibinfo {volume} {1}},\ \bibinfo
		{pages} {77} (\bibinfo {year} {1989})}\BibitemShut {NoStop}%
	\bibitem [{\citenamefont {Abid}\ and\ \citenamefont
		{Chesters}(1994)}]{Abid1994}%
	\BibitemOpen
	\bibfield  {author} {\bibinfo {author} {\bibfnamefont {S.}~\bibnamefont
			{Abid}}\ and\ \bibinfo {author} {\bibfnamefont {A.~K.}\ \bibnamefont
			{Chesters}},\ }\href {\doibase 10.1016/0301-9322(94)90033-7} {\bibfield
		{journal} {\bibinfo  {journal} {Int. J. Multiphase Flow}\ }\textbf {\bibinfo
			{volume} {20}},\ \bibinfo {pages} {613} (\bibinfo {year} {1994})}\BibitemShut
	{NoStop}%
	\bibitem [{\citenamefont {Klaseboer}\ \emph {et~al.}(2000)\citenamefont
		{Klaseboer}, \citenamefont {Chevaillier}, \citenamefont {Gourdon},\ and\
		\citenamefont {Masbernat}}]{klaseboer2000film}%
	\BibitemOpen
	\bibfield  {author} {\bibinfo {author} {\bibfnamefont {E.}~\bibnamefont
			{Klaseboer}}, \bibinfo {author} {\bibfnamefont {J.~P.}\ \bibnamefont
			{Chevaillier}}, \bibinfo {author} {\bibfnamefont {C.}~\bibnamefont
			{Gourdon}}, \ and\ \bibinfo {author} {\bibfnamefont {O.}~\bibnamefont
			{Masbernat}},\ }\href@noop {} {\bibfield  {journal} {\bibinfo  {journal} {J.
				Colloid Interface Sci.}\ }\textbf {\bibinfo {volume} {229}},\ \bibinfo
		{pages} {274} (\bibinfo {year} {2000})}\BibitemShut {NoStop}%
		\bibitem [{Sup()}]{Supp_mobile}%
	\BibitemOpen
	\href@noop {} {}\BibitemShut {NoStop}%
	\bibitem [{\citenamefont {Yaminsky}\ \emph
		{et~al.}(2010{\natexlab{b}})\citenamefont {Yaminsky}, \citenamefont
		{Ohnishi}, \citenamefont {Vogler},\ and\ \citenamefont
		{Horn}}]{Yaminsky2010c}%
	\BibitemOpen
	\bibfield  {author} {\bibinfo {author} {\bibfnamefont {V.~V.}\ \bibnamefont
			{Yaminsky}}, \bibinfo {author} {\bibfnamefont {S.}~\bibnamefont {Ohnishi}},
		\bibinfo {author} {\bibfnamefont {E.~A.}\ \bibnamefont {Vogler}}, \ and\
		\bibinfo {author} {\bibfnamefont {R.~G.}\ \bibnamefont {Horn}},\ }\href
	{\doibase 10.1021/la904482n} {\bibfield  {journal} {\bibinfo  {journal}
			{Langmuir}\ }\textbf {\bibinfo {volume} {26}},\ \bibinfo {pages} {8075}
		(\bibinfo {year} {2010}{\natexlab{b}})}\ \BibitemShut {NoStop}%
	\bibitem [{\citenamefont {Maali}\ \emph {et~al.}(2017)\citenamefont {Maali},
		\citenamefont {Boisgard}, \citenamefont {Chraibi}, \citenamefont {Zhang},
		\citenamefont {Kellay},\ and\ \citenamefont
		{W{\"u}rger}}]{maali2017viscoelastic}%
	\BibitemOpen
	\bibfield  {author} {\bibinfo {author} {\bibfnamefont {A.}~\bibnamefont
			{Maali}}, \bibinfo {author} {\bibfnamefont {R.}~\bibnamefont {Boisgard}},
		\bibinfo {author} {\bibfnamefont {H.}~\bibnamefont {Chraibi}}, \bibinfo
		{author} {\bibfnamefont {Z.}~\bibnamefont {Zhang}}, \bibinfo {author}
		{\bibfnamefont {H.}~\bibnamefont {Kellay}}, \ and\ \bibinfo {author}
		{\bibfnamefont {A.}~\bibnamefont {W{\"u}rger}},\ }\href@noop {} {\bibfield
		{journal} {\bibinfo  {journal} {Phys. Rev. Lett.}\ }\textbf {\bibinfo
			{volume} {118}},\ \bibinfo {pages} {084501} (\bibinfo {year}
		{2017})}\BibitemShut {NoStop}%
\bibitem [{\citenamefont {Katsir}\ and\ \citenamefont
	{Marmur}(2014{\natexlab{a}})}]{Katsir2014c}%
\BibitemOpen
\bibfield  {author} {\bibinfo {author} {\bibfnamefont {Y.}~\bibnamefont
		{Katsir}}\ and\ \bibinfo {author} {\bibfnamefont {A.}~\bibnamefont
		{Marmur}},\ }\href {\doibase 10.1021/la503373d} {\bibfield  {journal}
	{\bibinfo  {journal} {Langmuir}\ }\textbf {\bibinfo {volume} {30}},\ \bibinfo
	{pages} {13823} (\bibinfo {year} {2014}{\natexlab{a}})}\BibitemShut {NoStop}%
	\bibitem [{\citenamefont {Zhang}\ \emph {et~al.}(2017)\citenamefont {Zhang},
		\citenamefont {Manica}, \citenamefont {Tchoukov}, \citenamefont {Liu},\ and\
		\citenamefont {Xu}}]{zhang2017effect}%
	\BibitemOpen
	\bibfield  {author} {\bibinfo {author} {\bibfnamefont {X.}~\bibnamefont
			{Zhang}}, \bibinfo {author} {\bibfnamefont {R.}~\bibnamefont {Manica}},
		\bibinfo {author} {\bibfnamefont {P.}~\bibnamefont {Tchoukov}}, \bibinfo
		{author} {\bibfnamefont {Q.}~\bibnamefont {Liu}}, \ and\ \bibinfo {author}
		{\bibfnamefont {Z.}~\bibnamefont {Xu}},\ }\href@noop {} {\bibfield  {journal}
		{\bibinfo  {journal} {J. Phys. Chem. C.}\ }\textbf {\bibinfo {volume}
			{121}},\ \bibinfo {pages} {5573} (\bibinfo {year} {2017})}\BibitemShut
	{NoStop}%
	\bibitem [{\citenamefont {{Del Castillo}}\ \emph {et~al.}(2016)\citenamefont
		{{Del Castillo}}, \citenamefont {Ohnishi}, \citenamefont {Carnie},\ and\
		\citenamefont {Horn}}]{DelCastillo2016a}%
	\BibitemOpen
	\bibfield  {author} {\bibinfo {author} {\bibfnamefont {L.~A.}\ \bibnamefont
			{{Del Castillo}}}, \bibinfo {author} {\bibfnamefont {S.}~\bibnamefont
			{Ohnishi}}, \bibinfo {author} {\bibfnamefont {S.~L.}\ \bibnamefont {Carnie}},
		\ and\ \bibinfo {author} {\bibfnamefont {R.~G.}\ \bibnamefont {Horn}},\
	}\href {\doibase 10.1021/acs.langmuir.6b01949} {\bibfield  {journal}
		{\bibinfo  {journal} {Langmuir}\ }\textbf {\bibinfo {volume} {32}},\ \bibinfo
		{pages} {7671} (\bibinfo {year} {2016})}\BibitemShut {NoStop}%
	\bibitem [{\citenamefont {Chesters}\ and\ \citenamefont
		{Bazhlekov}(2000)}]{chesters2000effect}%
	\BibitemOpen
	\bibfield  {author} {\bibinfo {author} {\bibfnamefont {A.~K.}\ \bibnamefont
			{Chesters}}\ and\ \bibinfo {author} {\bibfnamefont {I.~B.}\ \bibnamefont
			{Bazhlekov}},\ }\href@noop {} {\bibfield  {journal} {\bibinfo  {journal} {J.
				Colloid Interface Sci.}\ }\textbf {\bibinfo {volume} {230}},\ \bibinfo
		{pages} {229} (\bibinfo {year} {2000})}\BibitemShut {NoStop}%
\end{thebibliography}

\newpage

\appendix
\centerline{\large{\bfseries Appendix}}
This appendix provides a brief introduction on the theoretical model. A formal supporting file is in preparation to cover the experimental method and more supporting experimental results. 
\section{Electrolyte transportation equation}

The electrolyte transportation equation (Eq.1 in manuscript) is rearranged using $\Gamma = k_1c$, where  $k_1=-\frac{1}{RT(1+\epsilon_{\pm})}\frac{d\sigma}{dc}$ is a constant value represent the property of electrolyte. Replacing $\Gamma$ with $k_1c$ is reasonable because of the fast partition of electrolyte between interfacial region and film solution in picoseconds, much faster than the experimental timescale that equilibrium is achieved instantaneously. In this scenario, $\Gamma$ relies on their local concentration $c$ in the film solution. 


Using  $\Gamma = k_1c$, $D_s=1.5D$, and after some algebra, we obtained the rearranged equation:

\begin{eqnarray}
\label{bulk_surface_step4}
\begin{aligned}
\frac{\partial C}{\partial t}=-\frac{h(U+\frac{h^2}{12\mu} \frac{\partial p}{\partial r})}{h+2k_1}\frac{\partial C}{\partial r} -\frac{2k_1}{(h+2k_1)}\frac{1}{r}\frac{\partial(rCU)}{\partial r}+ \frac{(h+3k_1)}{(h+2k_1)}\frac{D}{r}\frac{\partial}{\partial r}\left(r\frac{\partial C}{\partial r}\right) 
\end{aligned}
\end{eqnarray}

The above equation is coupled with the Young-Laplace equation for surface deformation~\cite{Chan2011a}, and the lubrication equation for liquid drainage~\cite{davis1989lubrication}.
\subsection{Young-Laplace equation}
\begin{equation}
\frac{\sigma}{2r}\frac{\partial}{\partial r}\left(r\frac{\partial h}{\partial r}\right)=\frac{2\sigma}{R_H}-(p+\Pi)
\end{equation}
where $\sigma$ is the air-water interfacial tension, and $2\sigma/{R_H}$ is the Laplace pressure (here, $R_H$ is the harmonic mean radius $R_H=2R_1R_2/(R_1+R_2)$ of the two bubbles). The disjoining pressure $\Pi$ typically has contributions from van de Waals force and electrical double layer force.  In this work, only the retarded van der Waals force ($\Pi_{vdw}=-A/(6\pi h^3)$) is considered that may rupture the film at a certain thickness~\cite{vakarelski2010dynamic,liu2019coalescence}, where $A$ is the Hamaker constant that varies with distance.

\section{Lubrication equation with mobile air-water interface}



The lubrication equation for liquid drainage is written as:
\begin{equation} \label{lub}
\frac{\partial h}{\partial t}=-\frac{1}{r}\frac{\partial}{\partial r}\left(rUh\right)+\frac{1}{12\mu_w r}\frac{\partial}{\partial r}\left(rh^3\frac{\partial p}{\partial r}\right)
\end{equation}

The first term on the right hand side of Eq.~\ref{lub} represents the flow induced by the mobile air-water interfaces featured by the interfacial velocity $U$ (see Fig. 1). The interfacial velocity $U$ is determined by the continuity of tangential shear stress across the interface ($\tau_b=\tau_f$), with the shear stress in de film given by

\begin{equation}\label{shear_stress}
\tau_f=\mu \frac{\partial u}{\partial z}|_{z=z^+} = -\frac{h}{2}\frac{\partial p}{\partial r}
\end{equation}
and the bubble air flow $\tau_b=\mu_{a}\partial u/\partial z|_{z=z^-}$. The second equality in Eq.~\ref{shear_stress} is obtained through derivation of the velocity profile of the flow inside the film, which is calculated from lubrication theory. Using the boundary integral method and the fact that information is only needed at the surface, the interfacial velocity is described as~\cite{abid1994drainage,davis1989lubrication}:
\begin{equation}\label{mobile_4}
U\left(r\right)=-\frac{1}{\mu_{a}}\int_{0}^{\infty}\Phi\left(r,\omega\right)\tau\left(\omega\right)d\omega
\end{equation}

\[\Phi\left(r,\omega\right)=\frac{k}{2\pi}\sqrt{\frac{\omega}{2r}}\int_{0}^{\pi}{\frac{\cos\alpha}{\sqrt{1-k^2\cos\alpha}}d\alpha}\]

\[k^2=\frac{2r\omega}{r^2+\omega^2}\]

The interfacial velocity $U$ depends on the viscosity ratio between water and air ($\mu_w/\mu_{a}$ $\sim$ 50), which is very large. Therefore, the film drainage velocity with mobile interfaces can be three or four orders of magnitude faster than that with immobile air-water interfaces ($U=0$). 

\section{Marangoni stress due to electrolyte concentration gradient}
The continuously change of electrolyte concentration within the thin liquid film can be predicted by solving the electrolyte transportation equation, lubrication equation, and Young-Laplace equation together. The concentration gradient, in radial direction, will induce the Marangoni stress that in turn affect the film thinning and deforming process. The change in surface tension appears to be linear over a wide range of electrolyte concentration. Using this information, the Marangoni stress can be calculated as
\begin{equation}
\frac{\partial \sigma}{\partial r}=\frac{\partial \sigma}{\partial C}\frac{\partial C}{\partial r}=k_2\frac{\partial C}{\partial r}
\end{equation}
where $k_2$ is approximated to be a constant that can be evaluated from experimental results. 

The Marangoni stress will modify the stress balance at the interface, and finally affect the surface velocity $U$ through Eq.~\ref{mobile_4}
\begin{equation}
\tau=\tau_s+\tau_m=-\frac{h}{2}\frac{\partial p}{\partial r}+\frac{\partial \sigma}{\partial r}
\end{equation}

\end{document}